\documentstyle[12pt]{article}
\def\gsim{\stackrel{>}{\sim}}
\def\lsim{\stackrel{<}{\sim}}
\def\beq{\begin{equation}}
\def\eeq{\end{equation}}

\begin{document}

\begin{centering}

{\large{\bf Gamma--Ray Bursters, Neutrinos, and Cosmology}}\\

\vspace{1cm}

T.~J.~Weiler\\

{\it Department of Physics \& Astronomy, Vanderbilt University,\\
Nashville, TN 37235}\\

\vspace{1cm}

W.~A.~Simmons, S. Pakvasa, \& J.~G.~Learned\\

{\it Department of Physics \& Astronomy, University of Hawaii, Manoa,\\
Honolulu HI 96822}\\

\vspace{1cm}

{\it preprint VAND--TH--94--20 \& UH--511--801--94}\\hep-ph/9411432\

{\it 11/26/94}\\
\end{centering}

\vspace{1cm}


\begin{abstract}

Gamma ray burst (GRB) objects are now widely thought to be at cosmological
distances, and thus represent enormous energy emission.  Gamma ray spectra
extending to $GeV$ energies suggest the possiblity of accompanying neutrino
emission, and there are several models proposed suggesting the potential
detectability of such coincident neutrino bursts.  With this in view, we
examine possible measurements that might be conducted to give experimental data
useful for astronomy, for cosmology and also neutrino properties.

Of interest to astronomy and cosmology, we show how measurement of neutrino
flavor ratios yields information on the nature and relative distance of the
source.   We point out that cosmological time dilation might be measured for
these sources using neutrinos, as has been done for photons, and  that neutrino
oscillation lengths in the range of $1$ to $10^5~Mpc$ can be probed with GRB
neutrinos.  We thus note that these sources may make possible the  first
non-electromagnetic measurements of the scale size of the universe.  We discuss
tests of the weak equivalence principle, tests for flavor dependent
gravitational couplings, and tests for long time scale variation of physical
constants.  We also show that a number of new bounds on neutrino properties
(charge, mass, speed, lifetime) could be facilitated to levels well beyond
those already inferred from the neutrino observation of SN1987A.

We also examine the implications of these physics opportunities for designers
of neutrino  telescopes.  We conclude that detection may be possible in planned
instruments if the spectra are power law extending to the $TeV$ energy region,
and if the neutrino fluxes are equal to or greater than the gamma ray fluxes.
We emphasize the importance of low energy detection in future experiments
for the tests described above.

\end{abstract}

\section{Introduction}

Recent observations of gamma ray bursts (GRBs) supports the idea that these
objects may be distributed over cosmological ranges\cite{fishman94}. The GRBs
may be among the most luminous objects in the universe, with peak gamma ray
luminosities $L\sim 10^{51}\,{\rm ergs}/\sec $.  There is evidence that the
range of intrinsic luminosities is very narrow, which might allow distance
determinations out to red--shifts of 2 or more\cite{horack94}.  It is plausible
that when such large amounts of energy are released in objects which are
evidently quite compact, pions will be produced in hadronic  collisions, and
that the GRBs may serve as standard candles in bursts of neutrinos as well. The
($\sim 1/E^2$) power law spectrum of gamma rays which is observed to extend at
least to several $GeV$, suggests the possibility of particle acceleration and
radiation from other than electromagnetic interactions of electrons. Paczynski
and Xu \cite{paczynski94} have recently proposed a specific GRB\ model in which
the gamma rays and neutrinos arise from decay of pions produced in shock front
collisions.  In a second class of models (e.g. Plaga \cite{plaga94}) gamma rays
are hypothesized to come from superconducting cosmic strings (SCSs); the
luminosities are very high and one expects neutrino emission as well.

Models of the first kind\cite{paczynski94} will have the following generic
features for the neutrinos emitted: the neutrino energies range over MeV to (a
few) GeV, or perhaps much higher, and since the source is $\pi $-decay, the
flavor content has the proportions $\nu _e:\nu _\mu :\nu _\tau =2:1:0$.  We
also note that due to gamma ray absorption within the source, the luminosity in
neutrinos can, in principle, exceed the observed luminosity in gamma rays,
potentially by a large factor\footnote{ Recent emphasis has centered on neutron
star models\cite{nemiroff94},  motivated in part by the  observation that the
gamma--ray energy released by GRBs at cosmological distances is of order of a
per cent of the binding energy of a neutron star if the emission is isotropic,
and less if the emission is beamed. In analogy to a supernova one might expect
99 to 99.9\% of the binding energy is expected to emerge as neutrinos, yielding
a neutrino--to--photon energy emission ratio of $10^2$ to $10^3$. }.   In
models of the second kind\cite{plaga94}, the neutrino flavor content is
expected to be $\nu_e: \nu_\mu:\nu_\tau \ = 1:1:1$ and the energies can reach
up to $10~TeV$.

In this paper we suppose that the GRBs are indeed cosmological standard candles
(at least in an ensemble average sense) for neutrinos and gammas, in order to
maximize the physical inferences to be drawn from coincident neutrino and
photon detection.   We investigate the opportunities for neutrino astronomy,
for cosmology, and for particle physics.  Some of this analysis may also be
applicable to other astrophysical neutrino sources, such as AGNs which are
expected to emit neutrinos over a very large energy range; AGNs will probably
have observational data exceeding what is possible from GRBs, but are unlikely
to have such short time pulse emission\footnote{ However, Markarian 421 has
been seen by the Whipple Observatory very--high--energy $\gamma$-ray telescope
to have a flux increase of a factor of ten on a time scale of one
day\cite{kerrick95}.}.  We hope that our analysis will be useful to designers
of neutrino telescopes and inspire them to consider GRB neutrino detection as
an important goal for future instruments.

\section{Neutrino Mass Mixing and Oscillations}

One of the unique properties of three generations of neutrinos, not shared with
photons, is that they may carry with them two additional scale lengths
associated with flavor oscillations. The oscillation lengths are related to the
neutrino mass differences according to

\begin{equation}
l_{ij}({\rm km})\simeq 2.5[\frac{p({\rm GeV})}{\delta m_{ij}^2({\rm
eV}^2)}],
\label{one}
\end{equation}

where i and j are flavors and $\delta m_{ij}$ is the, currently unknown,
mass difference between neutrinos of the relevant flavors. The probability of
observing neutrinos of a given flavor at any distance from their source is a
well known function of the distance, the oscillation length and mixing
parameters, as well as the initial conditions. Therefore, if the neutrinos are
produced in the same flavor ratio in each GRB, then the measured flavor ratios
can, in principle, determine the relative distances between various sources, at
least for the case of small mass differences and consequent long oscillation
lengths.   Absolute distances will not be given directly by the measured flavor
ratios, because the oscillation length itself will be difficult to measure
independently.

However, if GRBs are standard candles whose distances can be calibrated (e.g.,
by the time dilation of their spectra, or by gravity waveforms of inspiraling
objects\cite{schutz86}), then the measured flavor ratios may provide a direct
determination of very small neutrino mass differences. Competing models place
the GRBs (i) within the solar neighborhood ($D\lsim pc$), (ii) in the galactic
halo ($10^{-2} kpc \lsim D \lsim 10^2 kpc$), and (iii)  at cosmic distances
($D\gsim Mpc$) as we assume herein. Assuming detection of a correlated neutrino
with GeV energy, the $\delta m^2$'s which are probed at each distance scale are
$\gsim 10^{-13} eV^2$, $\gsim 10^{-14}$ to $\gsim 10^{-18} eV^2$, and $\gsim
10^{-19} eV^2$, respectively.

If the neutrino oscillation indications from the atmospheric neutrino
observations are correct ($\delta m^2_{\mu x}\sim 0.01~eV^2$, and $\theta_{\mu
x}$ large, $\sim 20^{\circ}-40^{\circ}$ \cite{fukuda94}), then the observed
flavor content of the neutrinos from pion--producing GRBs should show the same
effect, {\it viz.} a $\nu _\mu :\nu _e$ flavor content of 1.2:1 rather than
2:1.
This applies to models of the Paczynski-Xu kind. Furthermore, because the
atmospheric solution has a short oscillation length $l_{\mu x} \simeq 250
p({\rm GeV})$ km, this flavor ratio should be universal for all GRBs
independent of their distance; individual GRBs might have hot source spots
smaller in size than the  oscillation length $l$, but individual oscillations
will sum to zero in an ensemble average. For the models like that of Plaga,
based on SCS's, the democratic flavor mixture is unchanged by oscillations and
remains $\nu_\mu: \nu_e: \ \nu_\tau \ = 1:1:1$.  It may be thus possible to
distinguish between these two classes of source models.

If the atmospheric neutrino anomaly is not due to neutrino oscillations, then
there is a more interesting possibility for GRB neutrinos. According to the
see-saw mechanism \cite{yanagida79},

$$
m_\nu \sim m_D^2/M
$$

where $m_D$ is the generation--dependent Dirac mass of either the up quarks,
down quarks, or charged leptons $m_l$. If we use $m_D\sim m_l$ and $M\sim $\
Planck Mass, the neutrino masses\footnote {Neutrino mass differences induced by
scattering on the inter-galactic medium are expected to be smaller than these
numbers\cite{learned94a}.} are $m_{\nu _e}\sim 2\cdot 10^{-17}eV$,
$m_{\nu _\mu }\sim 10^{-12}eV$, $m_{\nu _\tau }\sim 3\cdot 10^{-10}eV$, and
hence $\delta m_{e\mu }^2\sim 10^{-24}{\rm eV}^2$ and $\delta m_{\mu \tau }^2
\sim 10^{-19}{\rm eV}^2$.

In this see--saw case, the oscillation lengths are in the cosmic range 1 to
$10^5$ Mpc for neutrino energies in the $GeV$ range\footnote{ For most
purposes in this paper, it is sufficient to equate time and distance as if the
universe were static.  The generalization to an expanding cosmology is reserved
for \S 4. }. There is a maximum distance over which a mixture of mass
eigenstates can be expected to remain coherent\cite{nussinov76}.  However, for
the tiny mass differences here, the coherence length appears to exceed $10^5$
Mpc.

The longer oscillation length possibilities exceed the size of the visible
universe $\sim H_0^{-1} = 3000 h^{-1}$ Mpc, where $H_0 = 100 h$ km/sec/Mpc is
the present value of the Hubble parameter. If $l$ exceeds $H_0^{-1}$, then the
neutrinos do not oscillate, and the flavor ratios observed at earth are just
those established at emission. On the other hand, for $l$ less than $\sim 10^4$
Mpc, the flavor content of the neutrinos from pion--producing GRBs will vary
strongly with distance; and again, for Plaga-like models the flux will remain
universal with no dependence on distance.

Alternatively, if mass differences are large (of order $0.01~eV^2$ or
larger), then the separation of the mass eigenstates in time offers
another handle on $\delta m^2$.  At a fixed energy E, the arrival times of the
$\nu$'s are separated by

$$
\delta t =
5\times 10^{-3} \frac{(L/100~Mpc)~(\delta m^2/10^{-2}~eV^2)}{(E/100~MeV)^2}
sec,
$$

assuming much smaller time difference at the source.  For low energies and
large $\delta m^2$, $\delta t$ can range from $10^{-3}~sec$ to several seconds.
With a spectrum of energies, the neutrino burst would be spread more than the
accompanying gamma ray burst.

\section{Other Neutrino Properties}

As with the observations of Supernova 1987A, the distance, short emission time
and trajectory through varying gravitational fields, leads to the potential for
some fundamental tests of neutrino properties, which are not possible in
terrestrial laboratories\cite{pakvasa90}.  In the following we enumerate some
of
the possibilities.

\subsection{Neutrino Lifetime}

 From the detection of neutrinos arriving form the $50kpc$ distant SN1987A it
was possible to place a limit on the (laboratory frame) lifetime of the
$\bar{\nu_e}$ of $\tau (\bar{\nu_e}) > 5 \cdot 10^{12}\ sec$ for $E_\nu$ of the
order of $10~MeV$.  In the same way, observation of  $\nu$'s from GRBs would
place bounds on  $\tau (\nu) > 10^{17} (D/Gpc)\ sec$, which for a $10~Mpc$
distant source is some $200$ times stronger than the bounds on $\bar{\nu_e}$
from SN1987A.  Of course this is really a bound on the lifetime of the dominant
mass eigenstate.

\subsection{Neutrino Electric Charge}

If neutrinos carry even a tiny electric charge (as favored in some theoretical
models\cite{ignatev94}), then the passage through magnetic field regions
enroute to the earth from a distant source creates an additional dispersion in
arrival time, $\delta t$.  From the observed upper bound on any additional
dispersion beyond the gamma pulse time distribution one can then derive a limit
on the neutrino charge from the Barbillieni-Cocconi
formula\cite{barbillieni87}:

$$
\frac{Q_\nu}{|e|} < \frac{\delta t}{D} \sqrt{\frac{<E>}{0.6 D B}}
\left(\delta E / E \right) ^{-1/2},
$$

where $\delta t$ is the burst duration,  $D$ is the flight distance, $B$ is the
root--mean--squared average magnetic field  experienced by the neutrinos
enroute, and $\delta E/E$ is the relative spread in energies.   For $\delta t
\sim 10^{-3}~sec$,  $D \sim 10~Mpc$, $<E> \sim 1~GeV$, $B \sim 10^{-12}$~Tesla
and $\delta E /E \sim 1$,  one has $Q_\nu /|e| < 10^{-27}$. Hence it may be
possible to improve considerably on the SN1987A limit of $Q_{\nu_e} < 10^{-14}
|e|$, and the somewhat better laboratory limit of $Q_{\nu_e} < 10^{-19} |e|$.
Moreover, one could place the first limit  on $\nu_\mu$, and possibly on
$\nu_\tau$.

\subsection{Neutrino Speed}

To the extent that the $\gamma$-ray and neutrino pulses coincide, new limits on
neutrino speed relative to the speed of light may be placed.  If
$\delta t = t_{\nu} - t_\gamma$, then

$$
1-v_{\nu}/c \le  \delta t / D .
$$

For $\delta t$ as large as 1 second this can place upper bounds of the order of
$10^{-15} (10 Mpc/D)$ on $1-\beta_{\nu}$,  to be compared with $10^{-9}$ (for
$\nu_e$) from SN1987A\cite{stodolsky88}.

\subsection{Equivalence principle}

The neutrinos from SN1987A were used to establish limits on parameters in
various theories of gravitation and to test whether the Weak Equivalence
Principle (WEP) is symmetric with respect to bosons and fermions, and with
respect to matter and anti-matter.  Specifically, the Shapiro delays for
gammas, neutrinos, and anti-neutrinos passing the Galactic nucleus were
compared and found to be the same, within errors \cite{longo88,pakvasa89}.
Also, SN1987A provided tests for the presence of proposed new forces in nature
and tests of the equality of parameters when applied to photons, neutrinos, or
anti-neutrinos \cite{pakvasa89,grifols94}. The same methods would apply to
neutrinos from GRBs. However, under our assumptions, the distances would be
greater and the impact parameter with the Galactic nucleus would vary from
event to event, offering much improved sensitivity and a new distance scale, as
well as a large increase in statistics.

To test the possibility that the neutrino and antineutrino gravity couplings
differ requires tagging of neutrino and antineutrino events separately.
Distinguishing between $\nu_e$ and $\bar{\nu_e}$ interactions at low energy may
be possible because of the charged current capture by protons.  At higher
energies distinguishing between $\nu_\mu$ and $\bar{\nu_\mu}$ interations is in
principle possible if one employed a magnetic field, or if one could adequately
detect the muon capture by nuclei; but in practice it cannot be done in
instruments proposed at this time.  A $\nu - \bar{\nu}$ separation would also
be useful for studying possible CP violation in neutrino
oscillations\cite{cabibbo78}.

A flavor violating gravitational coupling has been proposed as a possible
mechanism for accounting for atmospheric neutrino as well as  solar
anomalies\cite{pantaleone93}. It is remarkable that one choice of parameters
($sin^2 2 \theta_G \sim 1$, $\delta f \sim O(10^{-15})$) can account for both
problems.  The transition probability for $\nu_\mu \leftrightarrow \nu_e$ is
$sin^2(2\theta_G) sin^2 (\frac{1}{2} L E \phi (L) \delta f )$,  where $\phi$ is
the gravitational potential and $\delta f$ is a measure of the degree of
violation of the WEP.  Without a knowledge of $\phi$ along the path it is not
possible to calculate the net effect on the mixing, but it is very likely
similar to the expectation in the oscillation case ({\it i.e.}, $\nu_e:\nu_\mu$
becomes $1.2:1$ from $2:1$).

\section{Cosmology}

Norris, et al. \cite{norris94}, have analyzed the experimental data assuming
that the GRBs are standard candles in gamma rays and have found evidence for a
cosmological time dilation. If the GRBs are neutrino sources, then the same
analysis can provide, for the first time, non-electromagnetic evidence for the
expansion of the universe\footnote{ We note that such an analysis would exclude
any cosmological model which attempts to explain the observed red--shifts by
``tired photons''; earlier arguments against the tired photon hypothesis were
given by Zeldovich, {\it et al.}\cite{zeldovich63}. }.

The prevailing view, that the expansion of the universe is of a universal
nature, leads to the expectation that the gamma ray and neutrino time dilations
will be identical.  Unfortunately, a nearly identical dilation may occur if the
cause is evolutionary effects.  This is because the charged and neutral pions
of the pion--production model, and the emitted photons and neutrinos of the SCS
model arise from a common mechanism.   Still, it is possible that the neutrino
and photon opacities evolve differently, in which case studies of the differing
dilations may yield information on  cosmic evolution.

An important probe of the universe at an early time is  the oscillation phase
itself, $\phi_{ij}$. In Minkowski space, the quantum mechanical phase is just
$E t$. However, it is a bit more complicated in the expanding universe.  In the
adiabatic approximation, the phase is

\beq
\phi = R_0 \int_0^{\tau} \frac{d\tau'}{R(\tau')}E .
\label{phase}
\eeq

We have assumed a time--dependent Robertson--Walker metric,  with scale factor
$R(\tau)$; $\tau$ is the lookback time to the source emission, and $R(0)\equiv
R_0$. The red--shift factor $R_0/R(\tau)$ accounts for the time dilation of our
clocks compared to early universe clocks. Taking the difference of two
mass--eigenstate phases, expanding $E_i\simeq p + m_i^2/2p$, and red--shifting
the momentum to its  present--time observed value via $p(\tau) = (R_0/R(\tau))
p_0$, one obtains from Eq. (\ref{phase}) the very simple result for the
oscillation phase:

\beq
\phi_{ij}=\frac{\tau \delta m^2_{ij}}{2 p_0}.
\label{oscphase}
\eeq

The blue--shift of the inverse momentum exactly cancels the red--shift from
time dilation. Thus, the correct generalization of the oscillation phase from
Minkowski space to an expanding cosmology is obtained by replacing  laboratory
time with cosmic lookback time, or equivalently, replacing $l_{ij}$ in Eq.
(\ref{one}) with $c\tau_{ij}$.

This deceptively simple result is rich in cosmological information. In
particular, if $\delta m^2$ were {\it a priori} known,  then a measurement of
the flavor ratio and $E_{\nu}$ would directly yield $\tau$.  This is analogous
to obtaining the cosmic red--shift $z$ directly from a measurement  of a
photon's energy, when the unshifted spectral line of the photon source is {\it
a priori} known. Furthermore, $\tau$ contains as much information as is
conveyed by $z$. In fact, $\tau$, $z$, and the distance $D$ are  linearly
related for small $z, \tau, D$,  and nonlinearly related for large $z, \tau,
D$. For example, to first nonlinear order,  some Taylor series expansions
relating these three variables are\cite{kolb90}
$z(\tau)=H_0\tau+(1+\frac{q_0}{2})(H_0\tau)^2+\ldots$, and its inverse relation
$\tau(z)=H_0^{-1}\left[z-(1+\frac{q_0}{2})z^2+\ldots \right]$; and $D_L =
H_0^{-1} \left[z+\frac{1}{2}(1-q_0)z^2 +\ldots\right]$. Here, $D_L$ is the
``luminosity'' distance defined as  $D_L^2 = {\cal L}/4\pi{\cal F}$, where
${\cal L}$ is the absolute  luminosity at the source (energy/time), and ${\cal
F}$ is the fluence measured at earth (energy/time/area).  Results for other
distance definitions are similar; e. g. the ``proper'' distance ($R_0\times$
comoving cordinate distance),  is given as $D_P =  H_0^{-1}
\left[z-\frac{1}{2}(1+q_0)z^2 +\ldots\right]$. In all of these relations,
$q_0$ is the present value of the deceleration parameter.

The value of $q_0$ is unknown; since the GRBs seem to exist at cosmic
distances, it is possible that oscillation measurements with GRB neutrinos may
shed some ``neutrino light'' on this important parameter. An idependent
measurement of even two of the three variables $z, \tau$, and $D$ would
potentially determine $q_0$ and test cosmological models, because of  the
nonlinear relations.  The series expansions make it  clear that the linear
Hubble relation fails by a fractional  amount $z$ in red--shift, $H_0\tau$ in
lookback time, and $D/H_0^{-1}$ in distance. Interpreting the burst dilation of
fainter GRBs as due to time dilation leads to the estimate $z\sim$ 1 to 2 for
these GRBs, so there is indeed hope that oscillation measurements may yield a
value for $q_0$. We have seen that for small neutrino masses, it may be
possible to measure the phase of oscillations from pion--producing GRBs. Of
course, one would require a much better understanding of the structure and
mechanisms of GRBs than we have today, in order to draw useful conclusions.

Let us assume for the moment that such measurements can be made with enough
precision to yield values for $\tau \delta m^2_{ij}$. If an independent
measurement of $z$ or $D$ is available, then a single GRB would suffice to fix
$\delta m^2$, and a second GRB measurement would yield $q_0$. In this idealized
situation, we may also inquire about the  nature of higher order terms in the
nonlinear  expansions relating $z, \tau$, and $D$.  In fact, given a
cosmological model, the nonlinear relations among $z, \tau$, and $D$ are
exactly calculable.  For example,  $\tau = \int^{R_0}_{R(\tau)}dR/ \dot{R} =
\int^{R_0}_{R(\tau)} d\ln R/ H$ will yield $\tau(1+z=R_0/R(\tau))$ once
$\dot{R}(\tau)$ or equivalently,  $H\equiv \dot{R}/R$ are determined as a
function of $R$ (or $z$) from the Friedmann equation. Ignoring the radiation
energy density of the recent universe compared to the matter density, the
Friedmann equation reads
$H(z)^2 = H_0^2\left[(1+z)^2(1+z\Omega_0)-z(2+z)\Omega_{\Lambda}\right]$,
where $\Omega_0$ is the present matter density compared to the critical value
$\rho_{c}=3H_0^2/8\pi G$, and $\Omega_{\Lambda}=\Lambda/3H_0^2$; $\Lambda$ is
the comsological constant.   (The Friedmann universe is open or closed
according to whether  $\Omega_0 +\Omega_{\Lambda}$ is less than, or greater
than, unity.)   The Friedmann equation may be manipulated to yield

\beq
\tau=H_0^{-1}\int^{1+z}_1 \frac{d\, \omega}{\omega}
(\Omega_0 \omega^3 +\Omega_\Lambda -\Omega_k \omega^2)^{-\frac{1}{2}},
\label{tau}
\eeq

where the curvature term $\Omega_k = \Omega_0 +\Omega_{\Lambda} -1$  is a
constrained by the Friedmann equation itself. A thorough set of numerical
solutions to this equation  may be found in ref.\cite{felten86}. With
$\Omega_{\Lambda}$ omitted, the integral is easily solved analytically. The
form of the solution for $\tau(z)$ depends on whether the universe is closed
($\Omega_0 > 1$), critical ($\Omega_0 = 1$)  or open ($\Omega_0 < 1$). For the
critical case, motivated by inflation, the result is

\beq
\tau=\frac{2}{3} H_0^{-1}\left[1-(1+z)^{-3/2}\right].
\label{taucrit}
\eeq

For small $z$, the integral is just $z$, and the linear Hubble relation $H_0
\tau\simeq z$ results.  A related calculation yields the simple and useful
relation between the red--shift and the  luminosity distance\cite{weinberg72},
valid for $\Omega_{\Lambda} \ll \Omega_0$: $D_L =(2 H_0^{-1}/\Omega_0^2) \left[
z \Omega_0 + (\Omega_0 -2) (\sqrt{z\Omega_0 +1} -1) \right]$.  Then for
$\Omega_0 =1$ as suggested in inflationary cosmologies,  $D_L =
2H_0^{-1}\left[z-1-\sqrt{z+1}\right]$.  Given a cosmological model,
measurements of a second GRB would potentially validate or invalidate the
model,  by fitting or not fitting the nonlinear $\tau(z)$ or $\tau(D)$
relation.

In static Minkowski space, a  neutrino source is most useful for oscillation
studies if its distance is  comparable to the oscillation length; a shorter
distance does not provide sufficient path length for oscillations to develop,
and in a longer distance the information is effectively averaged over many
oscillations. However, in an expanding metric, distance must be carefully
defined, and the situation is different. We may calculate the lookback times
and luminosity distances to typical GRBs using the $\tau(z)$ or $D_L(z)$
formulae just given. The results are $\tau(z=1)=0.43 H_0^{-1}$ and
$\tau(z=2)=0.54 H_0^{-1}$, and $D_L(z=1)=1.18 H_0^{-1}$ and $D_L(z=2)= 2.54
H_0^{-1}$. Note that for this matter--dominated, critical--density example,
$D_L(z)$ is unbounded as $z$ increases, whereas the lookback time $\tau(z)$
has approaced $\frac{2}{3} H_0^{-1}$ for large $z$. The lookback  time is the
relevant variable for neutrino oscillation,  since it is directly proportional
to the oscillation phase, according to Eq. (\ref{oscphase}). The fact that
$\tau$ has become asymptotic for the red--shift  values typical of GRBs has an
important implication: the value of $\tau$ in the oscillation phase is nearly
$\tau(z=\infty)$, which in any cosmological model is a known fraction of
$H_0^{-1}$. This fact means that the uncertainty in $\tau$ is dominated by the
uncertainty in $H_0$, which is only a factor of two.   Thus, a single
measurement of the oscillation phase and neutrino energy may permit  a
determination of $\delta m^2$ to a factor of two!

Comparing an exact cosmological solution for $z, \tau$, or $D$  to the
appropriate Taylor series approximation relates each term in the series to more
fundamental quantities. The simplest relation is between the parameter $q_0$
multiplying the quadratic term, and the parameters $\Omega_0$ and
$\Omega_{\Lambda}$ in the Friedmann equation. The relation is\cite{felten86}
$q_0 = \Omega_0 /2 - \Omega_{\Lambda}$. Thus, a neutrino oscillation
determination of $q_0$ would establish a  fundamental constraint between the
two parameters of the Friedmann universe.

Studying cosmology by measuring neutrino flavor ratios has one tremendous
advantage over other methods, namely that flavor ratios should be independent
of any evolutionary effects in the ensemble of sources.  Absolute neutrino
luminosities may evolve, but the initial neutrino flavor ratios are fixed  by
microphysics; it is hard to imagine that these ratios  will change with cosmic
history. This is in sharp contrast to the use of distant ``candle''
luminosities to infer deviations from the linear Hubble Law, which may be due
to the deceleration of the universe or to evolutionary effects in the candles.

\section{Physical Constants - Time Dependence}

There is a long tradition in physics of asking whether various constants are,
in fact, constant over cosmological time. The best known of these questions was
raised by Dirac. In 1981, Barrow\cite{barrow81} reviewed these ideas. Among
his conclusions is the idea that only dimensionless constants can have a
meaningful time dependence.  In fact there are some new strong limits upon the
time variation of the fine structure constant and the electron to proton mass
ratio, and upon posible variations across causally disconnected regions of
space\cite{cowie94}.

However, any possible time variation of the dimensionless parameters upon which
neutrino oscillations depend (mass ratios and mixing angles, or equivalently,
the Yukawa couplings at the origin of fermion mass generation)  are without
constraint.  If the distances to the GRBs can be measured by independent means,
and the neutrino mass differences turn out to be very small, then a time
dependence of the mixing angles over cosmological times may be detectable as
deviations from the expected flavor ratios.

It has been speculated that dimensionful comological parameters may have a
time--dependence.  After all, the Hubble parameter itself has a complicated
dependence on time, varying inversely during power law expansion, and
exponentially during inflationary expansion.  The cosmological constant
$\Lambda(t)$ (or equivalently, the energy density of the vacuum), has been
hypothesized to relax to zero asymptotically with time \cite{antoniadis84}. And
the Hubble parameter and Newton's gravitational constant have been hypothesized
to consist of two terms each, one standard and one oscillating periodically in
time \cite{hill90}.  Motivation for the relaxation of $\Lambda$ is that the
present value of $\Lambda/8\pi G$ is known to be less than
$10^{-47}~GeV^4$\cite{kolb90}, whereas there is no good theoretical
understanding of why this should be so small. Motivation for the periodic
oscillation in $H$ and $G$ is that periodic modulation offers an explanation
for the controversial  observation\cite{broadhurst90} of a $128/h$ Mpc
quasi--period in deep ``pencil beam'' surveys of galaxy positions.  If
$\Lambda(\tau), G(\tau)$, and/or $H(\tau)$ are time--dependent, then the
Friedmann equation is modified. As discussed in the previous section, the
oscillation phases of neutrinos  emitted at large lookback time $\tau$ are
sensitive to the parameters in the (now modified) Friedmann equation.  Thus,
measured flavor ratios can offer information on the time dependences of these
cosmic parameters.

\section{Implications for Neutrino Telescopes}

We discuss some implications of our analysis for designers of neutrino
telescopes. The spectrum of gammas from GRBs has been seen out to energies of a
few $GeV$, so lacking a detailed GRB model one would do well to focus upon
neutrino detection in a similar energy range.   There are already weak limits
on the neutrino flux associated with GRBs from the IMB
experiment\cite{becker94} and others.  The largest deep mine experiment yet
planned, the SuperKamiokande detector with a  50 kiloton sensitive volume
(scheduled for operation in 1996) will have about ten times the sensitivity for
events with energies between $5~MeV$ and a few $GeV$ as had previous
instruments.

Much further progress for sensitivity down to the few $MeV$ region is not
presently on the horizon. One possibility would be parasitic use of  a megaton
size detector contructed for observation of supernova neutrinos out to a few
$Mpc$.  The capability to sense neutrino versus antineutrino interactions via
interaction characteristics, muon absorption, or magnetic fields to distinguish
charge, would provide a powerful tool for many of the tests described above.

However, if the GRB neutrino spectrum extends to energies of many $GeV$ or
even $TeV$, then there is more hope for the near future. This is because the
detectability of signals rises strongly with energy.   For example, the AMANDA,
Baikal, DUMAND and NESTOR ice/water instruments now  under construction have
effective volumes for $100~GeV$ neutrinos of order $10^6~tonnes$, up by two
orders of magnitude from underground instruments\cite{learned94b}.  Observation
of the ratio of muon charged current events to muonless events would
potentially allow for discimination between the putative neutrino--flavor
democratic source and the pion source with $\pi \rightarrow \mu \rightarrow e$
decay and $\nu_{\mu}$/$\nu_e$ = 2.

We can make a rough estimate of counting rates for neutrinos by assuming a
spectral shape which we take to be $1/E^2$, and a gamma flux which we take to
be $1~\gamma/cm^2/burst$ with energy greater than $1~MeV$.   The ratio of
neutrinos to gamma rays, labeled as $\eta$, could be zero in the case of a
purely electromagnetic origin of the gammas, in which case most of the
foregoing is irrelevant except for the important constraint upon the source
model.  For the situation of particle acceleration with power law spectra, as
discussed by Paczynski, $\eta \gsim 1$.   $\eta \simeq 1$ holds if the source
is not heavily shielded. Even better for our purposes, in situations similar to
that expected near AGN\cite{stenger92} the attenuation of the gamma rays can be
severe, while the neutrinos flow freely and so $\eta >> 1$, possibly even as
large as the $10^3$ energy emission ratio expected in supernovae.

We take a simplistic model of a standard $10,000~m^2$ effective area  (for muon
detection) instrument ({\it e.g.}, AMANDA, Baikal, DUMAND or NESTOR). In
underwater (or ice) experiments the area grows with energy somewhat, but for
simplicity we make the conservative assumption that this area independent of
energy (as in mine detectors).  We take the effective detector area for
neutrinos then as the the muon range, times the effective area for muons, times
the density of the medium, times Avogadro's Number, times the
neutrino--nucleon cross section.  This turns out to be $90~cm^2$ for the
nominal detector at $1~TeV$, and it scales roughly as ${E_\nu}^2$ from cross
section and range, for energies from $1~GeV$ to $10~TeV$.

A given detector will have some threshold detection energy, which in practice
is not a step function, though for simplicity we take it to be so, at $20~GeV$.
We also take an arbitrary maximum neutrino energy of $1~TeV$.   With these
assumptions we find for the expected number of neutrino interactions per burst
which are below the neutrino detector horizon, the value $9 \cdot 10^{-5} \eta
/ burst$ in muon neutrinos. If we take the rate of GRB (as now detected) as
about $1/day$, then the total expected number of correlated muons for this
standard neutrino detector  is $1.5 \cdot 10^{-2} \eta /year$.  Since the input
assumptions are certainly imprecise to a factor of ten, this could easily be
well detectable or beyond experimental reach.

If $\eta = 1$, it is easy to see why existing underground instruments have not
seen such correlations as yet, despite lower thresholds.  The IMB detector had
$400~m^2$ area, 25 times less than our assumed instrument.  In fact, one can
turn this around and ask what limit on $\eta$ is implied by the non--detections
in present underground instruments, under our assumptions of spectrum. In
Figure \ref{fig:eta}, we show the combinations of maximum GRB neutrino energy
and $\eta$ which would lead to one event detected per year in IMB ($400~m^2$),
a next generation instrument ($10^4~m^2$), and a hypothetical $km^3$ detector
(KM3).  One sees that for a $TeV$ GRB neutrino cutoff energy, the IMB limit
on the neutrino to gamma ratio is of the order of a few thousand.

Another approach is to consider the brightest GRBs, and ask whether or not
multiple events from a single source are possible.  If we assume that the
distribution of GRB gamma fluxes are dominated by spatial distribution, then
0.1\% of the GRBs will be at 0.1 of the distance of the typical burst, and
offer 100 times the neutrino flux at earth; this is the famous ``3/2 law''.
Thus in one year of operation during which there will be about 1000 GRBs, one
might find a GRB with $9 \cdot 10^{-3} \eta $ muons.  One can conclude that
detection of multiple muons per GRB is likely in the standard next generation
instrument only if $\eta \gg 1$.  This is illustrated in Figure
\ref{fig:brightest}, where the diagonal lines indicate the $\eta$ and maximum
neutrino energy combinations needed to see an event with ten muons once per
two years in the various classes of detectors (and roughly one in three such
GRBs would have a coincident GRO observation).

Better possibilities exist with a $km^3$ scale instrument.  Even though current
design discussions center around optimization for detection of $TeV$ to $PeV$
neutrinos from AGN's, neutrinos from GRB's (as well as other opportunities such
as dark matter searching and atmospheric neutrino oscillation measurement)
argue in favor of reducing the threshold to as low a value as feasible, say of
the order of $10~GeV$. Given the lack of understanding of the process which
generates the GRB's we cannot do much in terms of {\it a priori} optimization
of detector design.  Good timing is obviously important, but high angular
resolution is not as demanding as for neutrino point source searches.

If we take the energy threshold to again be $20~GeV$ for the ($km^3$) neutrino
detector with $10^6~m^2$ effective muon area, then under the same assumptions
we employed for the standard next generation instrument we get an expected
event  number of $9 \cdot 10^{-3} \eta /burst$, or about $1.5 \eta /year$ GRO
coincident detections.  More encouraging
yet is the rate of multiples: the brightest 0.1\% of GRBs might produce the
spectacular signature of $90 \mu$'s once in two years if $\eta$ should be 100
and the maximum energy $1~TeV$, in which case one could begin the studies
outlined above.

\section{Conclusion}

Any detection of GRBs in neutrinos would have great significance in
understanding these enigmatic objects.  What we advertise herein is that
moreover such a detection can lead to fundamental exploration of neutrino
physics, astrophysics,  and maybe even cosmology.  This exploration is not
possible by any other means we know.

The implications for the telescope designers then are fairly obvious: make
instruments with as low an energy threshold as practical, and allow for
upgrades (in terms of energy sensitivity and capability to resolve neutrino
flavors) to follow the path of discovery.

\section*{Acknowledgements}

We want to thank Xerxes Tata for many useful discussions.  We also acknowledge
Jack VanderVelde for a useful suggestion about calculating rates. We also want
to thank Andy Szentgyorgyi for help with the BATSE data.  This work was
supported in part by the U.S. Department of Energy grants no. DE-FG05-85ER40226
and no. DE-FG03-94ER40833.

\begin{figure}
\setlength{\unitlength}{0.012000in}%
\begingroup\makeatletter\ifx\SetFigFont\undefined
\def\x#1#2#3#4#5#6#7\relax{\def\x{#1#2#3#4#5#6}}%
\expandafter\x\fmtname xxxxxx\relax \def\y{splain}%
\ifx\x\y   
\gdef\SetFigFont#1#2#3{%
  \ifnum #1<17\tiny\else \ifnum #1<20\small\else
  \ifnum #1<24\normalsize\else \ifnum #1<29\large\else
  \ifnum #1<34\Large\else \ifnum #1<41\LARGE\else
     \huge\fi\fi\fi\fi\fi\fi
  \csname #3\endcsname}%
\else
\gdef\SetFigFont#1#2#3{\begingroup
  \count@#1\relax \ifnum 25<\count@\count@25\fi
  \def\x{\endgroup\@setsize\SetFigFont{#2pt}}%
  \expandafter\x
    \csname \romannumeral\the\count@ pt\expandafter\endcsname
    \csname @\romannumeral\the\count@ pt\endcsname
  \csname #3\endcsname}%
\fi
\fi\endgroup
\begin{picture}(467,520)(13,240)
\thicklines
\put( 70,360){\line( 1, 0){ 10}}
\put( 70,440){\line( 1, 0){ 10}}
\put( 70,520){\line( 1, 0){ 10}}
\put( 70,600){\line( 1, 0){ 10}}
\put( 70,680){\line( 1, 0){ 10}}
\put( 70,760){\line( 1, 0){ 10}}
\put( 80,360){\line( 0,-1){ 10}}
\put(160,360){\line( 0,-1){ 10}}
\put(240,360){\line( 0,-1){ 10}}
\put(320,360){\line( 0,-1){ 10}}
\put(400,360){\line( 0,-1){ 10}}
\put(480,360){\line( 0,-1){ 10}}
\put(465,360){\line(-1, 1){385}}
\put(305,360){\line(-1, 1){225}}
\put(480,485){\line(-1, 1){275}}
\put( 79,320){\makebox(0,0)[lb]{\smash{\SetFigFont{17}{20.4}{rm}1}}}
\put( 80,360){\framebox(400,400){}}
\put(154,320){\makebox(0,0)[lb]{\smash{\SetFigFont{17}{20.4}{rm}10}}}
\put( 26,460){\makebox(0,0)[lb]{\smash{
\SetFigFont{17}{20.4}{rm}Neutrino to
Gamma Ratio
}}}
\put(231,320){\makebox(0,0)[lb]{\smash{\SetFigFont{17}{20.4}{rm}100}}}
\put(317,320){\makebox(0,0)[lb]{\smash{\SetFigFont{17}{20.4}{rm}1}}}
\put(395,320){\makebox(0,0)[lb]{\smash{\SetFigFont{17}{20.4}{rm}10}}}
\put(472,320){\makebox(0,0)[lb]{\smash{\SetFigFont{17}{20.4}{rm}100}}}
\put(145,280){\makebox(0,0)[lb]{\smash{\SetFigFont{17}{20.4}{rm}GeV}}}
\put(385,280){\makebox(0,0)[lb]{\smash{\SetFigFont{17}{20.4}{rm}TeV}}}
\put( 60,360){\makebox(0,0)[lb]{\smash{
\SetFigFont{17}{20.4}{rm}1
}}}
\put( 60,435){\makebox(0,0)[lb]{\smash{
\SetFigFont{17}{20.4}{rm}10
}}}
\put( 60,510){\makebox(0,0)[lb]{\smash{
\SetFigFont{17}{20.4}{rm}100
}}}
\put( 60,585){\makebox(0,0)[lb]{\smash{
\SetFigFont{17}{20.4}{rm}1000
}}}
\put( 60,665){\makebox(0,0)[lb]{\smash{
\SetFigFont{17}{20.4}{rm}10000
}}}
\put(330,670){\makebox(0,0)[lb]{\smash{
\SetFigFont{17}{20.4}{rm}IMB
}}}
\put(160,535){\makebox(0,0)[lb]{\smash{
\SetFigFont{17}{20.4}{rm}KM3
}}}
\put(320,720){\makebox(0,0)[lb]{\smash{\SetFigFont{17}{20.4}{rm}Ruled Out By
IMB}}}
\put(250,610){\makebox(0,0)[lb]{\smash{
\SetFigFont{17}{20.4}{rm}BAND
}}}
\put(161,240){\makebox(0,0)[lb]{\smash{\SetFigFont{17}{20.4}{rm}Maximum GRB
Neutrino Energy}}}
\end{picture}
\caption{One event per year in coincidence with GRO for detectors
of $400~m^2$ (IMB), $10,000~m^2$ (BAND), and $10^6 ~m^2$ (KM3), for various
hypothetical neutrino to gamma ratios ($\eta$) and GRB maximum neutrino
energy.  The upper right hand corner region is already ruled out by IMB data.}
\label{fig:eta}
\end{figure}

\begin{figure}
\setlength{\unitlength}{0.012000in}%
\begingroup\makeatletter\ifx\SetFigFont\undefined
\def\x#1#2#3#4#5#6#7\relax{\def\x{#1#2#3#4#5#6}}%
\expandafter\x\fmtname xxxxxx\relax \def\y{splain}%
\ifx\x\y   
\gdef\SetFigFont#1#2#3{%
  \ifnum #1<17\tiny\else \ifnum #1<20\small\else
  \ifnum #1<24\normalsize\else \ifnum #1<29\large\else
  \ifnum #1<34\Large\else \ifnum #1<41\LARGE\else
     \huge\fi\fi\fi\fi\fi\fi
  \csname #3\endcsname}%
\else
\gdef\SetFigFont#1#2#3{\begingroup
  \count@#1\relax \ifnum 25<\count@\count@25\fi
  \def\x{\endgroup\@setsize\SetFigFont{#2pt}}%
  \expandafter\x
    \csname \romannumeral\the\count@ pt\expandafter\endcsname
    \csname @\romannumeral\the\count@ pt\endcsname
  \csname #3\endcsname}%
\fi
\fi\endgroup
\begin{picture}(467,520)(13,240)
\thicklines
\put( 70,360){\line( 1, 0){ 10}}
\put( 70,440){\line( 1, 0){ 10}}
\put( 70,520){\line( 1, 0){ 10}}
\put( 70,600){\line( 1, 0){ 10}}
\put( 70,680){\line( 1, 0){ 10}}
\put( 70,760){\line( 1, 0){ 10}}
\put( 80,360){\line( 0,-1){ 10}}
\put(160,360){\line( 0,-1){ 10}}
\put(240,360){\line( 0,-1){ 10}}
\put(320,360){\line( 0,-1){ 10}}
\put(400,360){\line( 0,-1){ 10}}
\put(480,360){\line( 0,-1){ 10}}
\put(480,360){\line(-1, 1){400}}
\put(320,360){\line(-1, 1){240}}
\put(480,475){\line(-1, 1){285}}
\put( 79,320){\makebox(0,0)[lb]{\smash{\SetFigFont{17}{20.4}{rm}1}}}
\put( 80,360){\framebox(400,400){}}
\put(154,320){\makebox(0,0)[lb]{\smash{\SetFigFont{17}{20.4}{rm}10}}}
\put(320,710){\makebox(0,0)[lb]{\smash{\SetFigFont{17}{20.4}{rm}Ruled Out By
IMB}}}
\put(231,320){\makebox(0,0)[lb]{\smash{\SetFigFont{17}{20.4}{rm}100}}}
\put(317,320){\makebox(0,0)[lb]{\smash{\SetFigFont{17}{20.4}{rm}1}}}
\put(395,320){\makebox(0,0)[lb]{\smash{\SetFigFont{17}{20.4}{rm}10}}}
\put(472,320){\makebox(0,0)[lb]{\smash{\SetFigFont{17}{20.4}{rm}100}}}
\put(145,280){\makebox(0,0)[lb]{\smash{\SetFigFont{17}{20.4}{rm}GeV}}}
\put(385,280){\makebox(0,0)[lb]{\smash{\SetFigFont{17}{20.4}{rm}TeV}}}
\put( 60,360){\makebox(0,0)[lb]{\smash{
\SetFigFont{17}{20.4}{rm}1
}}}
\put( 60,435){\makebox(0,0)[lb]{\smash{
\SetFigFont{17}{20.4}{rm}10
}}}
\put( 60,510){\makebox(0,0)[lb]{\smash{
\SetFigFont{17}{20.4}{rm}100
}}}
\put( 60,585){\makebox(0,0)[lb]{\smash{
\SetFigFont{17}{20.4}{rm}1000
}}}
\put( 60,665){\makebox(0,0)[lb]{\smash{
\SetFigFont{17}{20.4}{rm}10000
}}}
\put(160,535){\makebox(0,0)[lb]{\smash{
\SetFigFont{17}{20.4}{rm}KM3
}}}
\put(250,610){\makebox(0,0)[lb]{\smash{
\SetFigFont{17}{20.4}{rm}BAND
}}}
\put(161,240){\makebox(0,0)[lb]{\smash{\SetFigFont{17}{20.4}{rm}Maximum GRB
Neutrino Energy}}}
\put( 26,460){\makebox(0,0)[lb]{\smash{
\SetFigFont{17}{20.4}{rm}Neutrino to
Gamma Ratio
}}}
\put(325,665){\makebox(0,0)[lb]{\smash{
\SetFigFont{17}{20.4}{rm}IMB
}}}
\end{picture}
\caption{10 muons in one burst for brightest GRBs, seen in detectors
of $400~m^2$ (IMB), $10,000~m^2$ (BAND), and $10^6 ~m^2$ (KM3), for various
hypothetical neutrino to gamma ratios ($\eta$) and GRB maximum neutrino
energy.  The upper right hand corner region is already ruled out by IMB data.
The rate may be 0.5/year.}
\label{fig:brightest}
\end{figure}

\end{document}